\documentstyle[amssymb,preprint,aps]{revtex}

\begin{document}
\title{Acoustic and relaxation processes in supercooled o-ter-phenyl by
optical-heterodyne transient grating experiment}
\author{R. Torre$^{1,2}$, A. Taschin$^{1,3}$, M. Sampoli$^{1,4}$.}
\address{$^{1}$ INFM, Unit\`{a} di Firenze, largo E.Fermi 2, I-50125, Firenze, Italy\\
$^{2}$LENS, Universit\`{a} di Firenze, largo E.Fermi 2, I-50125,Firenze,Italy\\
$^{3}$Dip. di Fisica, Universit\`{a} di Firenze, largo E.Fermi 2, I-50125,Firenze, Italy.\\
$^{4}$Dip. di Energetica, Universit\`{a} di Firenze, via S.Marta, Firenze,
Italy.}
\date{6/1/2001}
\maketitle

\begin{abstract}
The dynamics of the fragile glass-forming o-ter-phenyl is investigated by
time-resolved transient grating experiment with an heterodyne detection
technique in a wide temperature range. We investigated the dynamics
processes of this glass-former over more then 6 decades in time with an
excellent signal/noise. Acoustic, structural and thermal relaxations have
been clearly identify and measured in a time-frequency window not covered by
previous spectroscopic investigations. A detailed comparison with the
density response function, calculated on the basis of generalized
hydrodynamics model, has been worked out. 
\end{abstract}

\pacs{64.70.PF, 78.20Jq, 78.47.+p}

\section{Introduction}

Supercooled liquids and glasses have been the subject, in these last years,
of an extensive theoretical and experimental study \cite{Books,Gotze99}. One
of most important features of these materials is the wide range of times
scales over which relaxations occur at fixed and mostly changing
temperature. The relaxation times can span over many decades (up to 13
decades) when temperature is varied from the melting to glass transition
temperature. Indeed these relaxation processes have a complex nature that,
despite the recent research efforts, remain unclear in many aspects. Among
the others, a fundamental issue is the interplay of different dynamical
variables present in these materials. So that the rotational dynamics and
its coupling with the translational variables are under\ an intense
experimental investigation \cite{OKE,Taschin,DIHARD}. Even when a single
variable (e.g. density) is studied, many processes are involved in the glass
dynamics (like for example the acoustic, structural and thermal relaxation)
and they are usually characterized by different time scales. From the
experimental point of view, it is of fundamental importance to measure the
investigated variables over the full temporal or frequency range. Often this
is accomplished hardly by \ the combination of several techniques (e.g.
Light Scattering, Photon Correlation Spectroscopy, etc.) since, even if the
same experimental observable is tested, the very wide frequency or time
domain with possible not overlapping range, preserves a reliable analysis of
the data.

The transient grating technique (TG) is well know to give a unique
experimental insight in the dynamics of fluid and glassy material \cite
{Eichler86}. Recently this technique has been applied to the study of
dynamics of supercooled liquids \cite{Nelson}. Using a CW beam probe it has
been covered a very wide temporal range, from $n\sec $ to $m\sec $, in a
single experiment giving an unique access to the glass dynamics.

In this paper we present a study of relaxation dynamics of the fragile
glass-former ortho-ter-phenyl (OTP) by optical Heterodyne Detected Transient
Grating (HD-TG) experiment. Acoustic, thermal and structural relaxations are
investigated in wide temperature range, from above the melting down to the
glass-transition. This HD-TG experiment produces data with an extremely high
signal/noise ratio and allows to detected and measure signals even of very
low level. We present a study of the complex relaxation pattern of the
density in a prototype of the fragile glass-formers, OTP, and we put in
evidence \ some new features of the slow thermal relaxation that can not be
properly account for by the standard hydrodynamic model.

The paper begins (sec. II) with a brief description of the transient grating
experiment and in particular underlines the improvements that can be
obtained by using an optical heterodyne detection. In sec. III the
theoretical background needed for data analysis and interpretation is given.
A detailed description of the OTP measurements is written in sec. IV and in
sec. V we present all data and their analysis. The last sec. VI is devoted
to discussions and conclusions.

\section{Heterodyne Transient Grating Experiment}

In a transient grating (TG) experiment two high power laser pulses interfere
inside the sample and produce a spatially periodic variation of the index of
refraction\cite{Eichler86,Yan87}. A third laser beam, typically of different
wavelength, is acting as a probe by impinging on the induced grating at the
Bragg angle and producing a diffracted beam, spatially separated by the pump
pulses and probe beam, see fig.\ref{set-up}. This diffracted beam is the
signal measured in a TG experiments and yields the dynamic information from
the relaxing TG. The spatial modulation in the TG defines a wave-vector $q$
characterizing the diffracted beam and hence the signal. The $q$ wave-vector
is:

\begin{equation}
q=\frac{4\pi \sin \left( \frac{\theta _{e}}{2}\right) }{\lambda _{e}}
\label{wavevector}
\end{equation}

where $\lambda _{e}$ and $\theta _{e}$ are the wavelength and the incidence
angle of the excitation laser pulses. When, as usual, the homodyne scheme is
used to detect the diffracted beam, the diffraction efficiency results to be
proportional to the square of the refraction index variation so that small
variations produce even smaller signal. A considerable improvement can be
obtained by using an optical heterodyne detection (HD). Indeed in our HD-TG
experiment, the measured signal is given by:

\begin{equation}
S(q,t)\propto \left\langle \left| E_{d}(q,t)\right| ^{2}\right\rangle
+\left\langle \left| E_{l}\right| ^{2}\right\rangle +2\left\langle \left|
E_{d}(q,t)\right| \right\rangle \left\langle \left| E_{l}\right|
\right\rangle \cos \Delta \varphi  \label{Signal}
\end{equation}

where $E_{d}$ is the electric field diffracted by the TG, $E_{l}$ is the
local beating field and $\Delta \varphi $ is the phase difference between
the $E_{d}$ and $E_{l}$, see fig.\ref{set-up}, $\left\langle \cdot
\right\rangle $ represent the time-averaging over the optical period, e.g. $%
\left\langle \left| E_{d}(q,t)\right| \right\rangle $ is the amplitude of
the oscillating diffracted electric field. The first, second and third terms
in the r.h.s. of eq. \ref{Signal}, are the homodyne ($\left\langle \left|
E_{d}\right| ^{2}\right\rangle $), local field ($\left\langle \left|
E_{l}\right| ^{2}\right\rangle $), and heterodyne contributions ($%
2\left\langle \left| E_{d}(q,t)\right| \right\rangle \left\langle \left|
E_{l}\right| \right\rangle \cos \Delta \varphi $) respectively. Normally,
the local field is kept constant and much higher than the diffracted one, so
that the homodyne contribution becomes negligible and the time variation of
the signal is dominated by the heterodyne term. In addition, this last term
can be\ experimentally isolated by subtracting two recorded signals with
different phases, and in particular the first one, $S_{+}$, with $\Delta
\varphi _{+}=2n\pi $ and the second, $S_{-}$, with $\Delta \varphi
_{-}=(2n+1)\pi $ ($n$ integer). It is evident that:

\begin{equation}
S_{HD}(q,t)=\left[ S_{+}-S_{-}\right] \propto \left\langle \left|
E_{d}(q,t)\right| \right\rangle \left\langle \left| E_{l}\right|
\right\rangle   \label{SignalHD1}
\end{equation}
We have two major advantages in using the heterodyne instead of homodyne
detection. First, we can improve substantially the signal to noise ratio in
the all time window, both because of the high level of the local field and
because of discarding spurious signals which are not reversed by a $\pi $
phase shift of the local field. Second, the sensitivity at long times when
the TG is vanishing, is enhanced enormously since the recorded signal is
proportional to $\left\langle \left| E_{d}\right| \right\rangle $ instead of
being proportional to its square. In the studies of materials with a weak
scattering efficiency and complex response, these features turns out to be
of basic importance. Nevertheless, the effective realization of such a
detection is quite difficult at optical frequencies. Indeed to get an
interferometric phase stability between the diffracted and the local field
is not a simply experimental task, and so only few HD-TG experiments have
been realized up to now \cite{Eichler86}. Recently the introduction of phase
gratings in the optical setup of TG experiments has reduced considerably the
difficulties of using a heterodyne detection \cite{Maznev98,Goodno98}. The
details of this new setup will be described in Sec. IV A.

Formally, in the limit of linear response theory the average diffracted
field is proportional to the specific response function of the material
convoluted with the excitation force produced by the laser excitation\cite
{Yan87}. Assuming the impulsive limit in time (i.e. the excitation time much
shorter than the observable characteristic times) and in wave-vector (i.e.
the excitation spot size much larger than the material wavelength scale) we
have:

\begin{equation}
S_{HD}(q,t)\propto \left\langle \left| E_{d}(q,t)\right| \right\rangle
\propto R(q,t).  \label{SignalHD2}
\end{equation}

The response function, $R(q,t)$, describes how the exciting pulses are
effective in producing a TG, i.e a periodic variation of the optical
properties out of equilibrium, and how this variation is relaxing toward
equilibrium. If the duration of exciting pulses has to be taken into
account, i.e. if some characteristic times in the responses can not be
considered long enough with respect to the pulse duration, the total
response has to be calculated from the appropriate convolution.

The response function has a tensorial nature, $R_{ijkl}$, where the
different component are selected through the excitation, probing and
detection directions of polarizations. In the present experiment all these
polarizations are taken vertical, i.e. normal to the scattering plane, (see
Sec.IV A for details) and then we are dealing with a single tensorial
component of the response function, $R_{VVVV}$.

The interaction of the excitation laser field with the material is
responsible of generating the TG and defining the response function.
Depending on the nature of this interaction, the excitation laser pulses can
produce a modulation of the real part of the refraction index (i.e. a
birefringence and/or a phase grating), and/or a modulation of the imaginary
part (i.e a dichroic and/or absorption grating). Which grating is excited
and how its dynamics reflects the material properties is a complex problem
that can be solved only under certain conditions and approximations.

As well as other molecules of glass forming materials investigated by TG 
\cite{Nelson}, our OTP molecule has no electronic absorption band at the
pump and probe wavelengths. However the strong near infrared pulses are
absorbed weakly by overtones and/or combinations of vibrational bands.
Typically, these vibrational excitations thermalize in few picoseconds.
According to Nelson and co-worker \cite{Nelson} the induced TG can be
described approximately as a pure phase grating generated by two different
mechanisms of laser-matter interaction: a temperature grating produced by
the field induced heating and a pressure electrostrictive grating due to the
field gradient. In this approach, no birefringence effects are taken in
account, such as a modification of the molecular polarizability orientation
due to field effects (optical Kerr effect) or to a molecular alignment
induced by roto-translation coupling. For materials composed by molecules of
nearly isotropic shape and without strong anisotropic interactions, such as
the OTP under the present investigation, the total response should be
considered mainly driven by the density. We verified experimentally this
hypothesis measuring the $S_{HHVV}$ signal (with the probe and detection
polarizations, $HH$, perpendicular to the excitation one, $VV$) and
comparing that with the $S_{VVVV}$ signal: no meaningful difference has been
detected for any temperature in OTP. So in this molecule no relevant
birefringence effects are present confirming that the response function can
be mainly described by the density dynamics. Viceversa strong difference has
been detected in a other glass-former, (e.g. {\it m}-toluidine and salol),
showing that the response function of glass-formers made of anisotropic
molecules must include birefringence and roto-translational coupling effects
\cite{Taschin}.

By taking into account the two mechanism above mentioned, the measured
signal is proportional to the q-component of the density variations and is
composed of three different contributions: roughly speaking, two coming from
the heating and one from the electrostriction. i) The part of the solid like
or fast liquid response of the sample to the sudden heat flux, launches a
sound wave of frequency $\omega _{A}=c_{A}\,q$ around the mean density at
the new local temperature. The sound wave is damped with the acoustic time
constant $\tau _{A}$ and then the local temperature relaxes with a time
constant $\tau _{H}$ due to the thermal diffusivity. Therefore this
contribution can be written as: $\rho _{q}^{(h,s)}\varpropto \left[ 1-\exp
(-t/\tau _{A})\cos (\omega _{A}\,t)\right] \exp (-t/\tau _{H}).$ ii) The
part of viscous liquid response to the heat flux produces a slow density
variation which grows up with a time constant $\tau _{R}$ and vanishes again
with the thermal diffusivity constant, i.e. $\rho _{q}^{(h,r)}\varpropto %
\left[ 1-\exp (-t/\tau _{R})\right] \exp (-t/\tau _{H}).$

iii) The sudden density momentum change due to the electrostriction launches
a sound wave of the same frequency $\omega _{A}=c_{A}\,q$ around the normal
density which is damped always with the same acoustic time constant, i.e. $%
\rho _{q}^{(e)}\varpropto \exp (-t/\tau _{A})\sin (\omega _{A}\,t).$

Intuitively the sum of these contributions will be the measured signal.
Obviously there can be interactions among the contributions if the time
constants are not well separated, indeed this separation has been supposed
implicitly in the previous derivation. Further, the viscous response in
supercooled liquids is made of a distribution of time constants instead of
single time. In the next section we derive the forced density fluctuations
from the hydrodynamic equations and important relations among the parameters
we have introduced.

\section{Theoretical Background}

The generalized hydrodynamics equations have been often used to describe the
frequency resolved light scattering experiments \cite{Pecora76}. They
express the conservation law the dynamics of the fluid is subjected to: the
conservation of mass, momentum and energy. As usual, we neglect the effects
of rotational dynamics, as already discussed and we use the linearized form
of the hydrodynamic equations (small fluctuation out of equilibrium)\cite
{Boon80}. Since we are interested to the q-component of the density
fluctuations, we write directly the Fourier-Laplace transform (FLT) of the
continuity, Navier-Stokes, and energy equations. They read 
\[
s\ \widetilde{\delta \rho _{q}}(s)+i\ q\ \widetilde{J_{q\parallel }}%
(s)=\delta \rho _{q}(0)
\]
\[
\lbrack s+\phi _{L}(s)\ q^{2}]\ \widetilde{J_{q\parallel }}(s)+i\
(qc_{o}^{2}/\gamma )\ [\widetilde{\delta \rho _{q}}(s)+\alpha \ \rho _{o\ }%
\widetilde{\delta T_{q}}(s)]=J_{q\parallel }(0)
\]
\[
(s+\gamma \chi q^{2})\ \rho _{o\ }\widetilde{\delta T_{q}}(s)+i\ [q(\gamma
-1)/\alpha ]\widetilde{\ J_{q\parallel }}(s)\delta \rho _{q}(s)+i\ q\
J_{q\parallel }(s)=\rho _{o\ }\delta T_{q}(0)
\]

where $\widetilde{\delta x_{q}}(s)$ stand for the FLT of the fluctuating
quantity$\ \delta x({\bf r},t)$, i.e. 
\[
\widetilde{\delta x_{q}}(s)=\int_{0}^{\infty }dt\ \exp (-st)\delta
x_{q}(t)=\int_{0}^{\infty }dt\ \exp (-st)\int_{-\infty }^{+\infty }d^{3}r\
\exp (i{\bf q\cdot r})\ \delta x({\bf r},t) 
\]

and $\rho _{o\ }$is the equilibrium mass density, $J_{q\parallel }$ the
component parallel to ${\bf q}$ of the mass current density, $\delta T_{q}$
the fluctuating temperature, $c_{o}$ the adiabatic zero frequency sound
velocity, $\gamma =C_{p}/C_{v}$\ the ratio of the specific heats, $\alpha
=-(\partial \rho /\partial T)_{p}/\rho $\ the coefficient of thermal
expansion, \ $\chi ={\em k}(\rho _{o}C_{p})^{-1}$\ the thermal diffusivity,
and $\rho _{o\ }\phi _{L}$\ the longitudinal kinematic viscosity.

Following Nelson\cite{Nelson,Yang95}, we can solve formally the previous
equations with the starting conditions $\delta \rho _{q}(0)=0$, $%
J_{q\parallel }(0)=i\,q\,F$, and $\rho _{o\ }\delta T_{q}(0)=Q/C_{v}$ and we
obtain that the HD signal can be written in the Nelson's notation as:
\begin{equation}
S_{HD}(q,t)\propto R_{\rho }(q,t)=-FG_{\rho \rho }(q,t)+QG_{\rho T}(q,t)
\label{Response}
\end{equation}

where $G_{\rho \rho }(q,t)$ is the response to the electrostriction (also
called Impulsive Stimulated Brilluoin Scattering, ISBS) , and $G_{\rho
T}(q,t)$ to the absorption (also called Impulsive Stimulated Thermal
Scattering, ISTS)\cite{Eichler86,Nelson}. $F$ and $Q$ are two constants that
define, respectively, the magnitude of the electrostriction and heat flux in
the limit of infinitely short pump laser pulses.

In order to calculate the density response function, we have to know all the
coefficients and in particular the time (or frequency) dependence of the
kinematic viscosity $\phi _{L}$. Only for particular form of $\phi _{L}$ it
is possible to obtain analytical expressions for $G_{\rho \rho }(q,t)$ and $%
G_{\rho T}(q,t)$. One of them is the very simple but useful Debye model of a
viscous fluid where the kinematic viscosity (or memory function for the
elastic modulus) is approximated by the sum of an instantaneous plus a
exponential relaxing term: $\phi _{L}(q,t)=v_{L}\delta (t)+(c_{\infty
}^{2}-c_{0}^{2})\exp (-t/\tau _{R}).$ If all the characteristic times (i.e.
the period of the acoustic wave, its damping time constant, the structural
relaxation time and the thermal diffusion time constant) are well separated
each other \ , very simple equations are obtained \cite
{Nelson,Yang95,ISBSstructural}: 
\begin{equation}
G_{\rho T}(q,t)\simeq A\left[ e^{-\Gamma _{H}\,t}-e^{-\Gamma _{A}\,t}\cos
\left( \omega _{A}t\right) \right] +B\left[ e^{-\Gamma _{H}\,t}-e^{-t/\tau
_{R}^{\prime }}\right]   \label{ResponseISTS}
\end{equation}
\begin{equation}
G_{\rho \rho }(q,t)\simeq C\left[ e^{-\Gamma _{A}t}\sin \left( \omega
_{A}t\right) \right]   \label{ResponseISBS}
\end{equation}

where: $\Gamma _{H}$ is the thermal damping rate ($\Gamma _{H}=1/\tau
_{H}=\chi q^{2})$, $\tau _{R}^{\prime }$ the effective structural relaxation
time ($\tau _{R}^{\prime }=\left( \frac{c_{A}}{c_{0}}\right) ^{2}\tau _{R}$
being $c_{A}$ the sound velocity), $\omega _{A}$ and $\Gamma _{A}=1/\tau _{A}
$ are the frequency and damping of acoustic longitudinal phonon, $A$, $B$,
and $C$ are constant dependent on sample and experimental setup. The
importance of this treatment resides in the possibility to extract other
information and to make expectations on the general behavior of the various
parameters versus $q$ and temperature. In particular we expect a maximum on
the acoustic damping both versus $q$ and $T$, and a gradual shift of the
sound velocity from $c_{0}$ to $c_{\infty }$ lowering the temperature or
increasing $q$. We have\cite{Yang95}: 
\begin{equation}
\omega _{A}=c_{A}q=\left[ c_{0}\sqrt{D+\sqrt{D^{2}+\left( c_{0}q\tau
_{R}\right) ^{-2}}}\right] q;\qquad D=[c_{\infty }^{2}/c_{0}^{2}-\left(
c_{0}q\tau _{R}\right) ^{-2}]/2
\end{equation}
\begin{equation}
\frac{\Gamma _{A}}{q^{2}}=\frac{1}{2}\left\{ \left[ v_{L}+\chi \left( \gamma
-c_{0}^{2}/c_{\infty }^{2}\right) \right] +\frac{c_{\infty }^{2}-c_{0}^{2}}{%
1+\omega _{A}^{2}\tau _{R}^{2}}\right\}   \label{gamma_q}
\end{equation}

Further, always in the assumption that $\omega _{A}\gg \Gamma _{A}\gg \tau
_{R}^{-1}\gg \Gamma _{H}$ and some other minor approximations\cite{Yang95},
we can extract the Debye-Waller factor or non-ergodicity parameter $%
f_{q\rightarrow 0}(T)$ from $A$ and $B$ coefficients which in turn can be
derived by fitting the experimental data with the theoretical expression:
\begin{equation}
f_{q\rightarrow 0}(T)=1-\frac{c_{0}^{2}}{c_{\infty }^{2}}=\frac{B}{A+B}
\label{DebeyWaller}
\end{equation}

However, the Debey model for the longitudinal kinematic viscosity has severe
limitations. As it has been pointed out in many experimental works, the
structural relaxation around the critical temperature of the structural
arrest is not reproduced by a single relaxation time but by a distribution
of them; often this fact has been attributed to the rise of heterogeneities
during the cooling. As a consequence, more complex models have to be used to
take into account thedistribution of \ relaxation times, e.g. by a
multi-exponential or stretched exponential approach \cite{Books}. When all
the relaxation times of \ the actual distribution are separated from the
other characteristic times, the kinematic viscosity affect only the term
with the $B$ coefficient (the ii) term of the previous section) and a
complex relaxation can be put directly in the response as a stretched
exponential, i.e. a Kohlrausch-Williams-Watts (KWW) function\cite{Yang95}:

\begin{equation}
G_{\rho T}(q,t)\simeq A\left[ e^{-\Gamma _{H}\,t}-e^{-\Gamma _{A}\,t}\cos
\left( \omega _{A}t\right) \right] +B\left[ e^{-\Gamma _{H}\,t}-e^{-\left(
t/\tau _{S}\right) ^{\beta }}\right]   \label{Responsebeta}
\end{equation}
where the 'relaxation time' $\tau _{S}$ and the stretching factor $\beta $
are fitting parameters for the relaxation time distribution. This
approximation has the advantage to keep an easy and readable form for the
response function but this simplicity prevents to account for possible
interactions among the different relaxing mechanisms. To compare $\tau _{S}$
with the relaxation times derived by other time distribution (like the
Cole-Davidson in the frequency domain) or derived in single relaxation time
approximation, the mean of the time distribution is calculated as $%
\left\langle \tau _{S}\right\rangle =\beta ^{-1}\Gamma (\beta ^{-1})\,\tau
_{S}$.

We want to remark that the presence of a stretched exponential in the $%
G_{\rho T}(q,t)$ can not be considered {\it per se} a way to introduce the
mode-coupling theory in the hydrodynamic response\cite{Goetze92}.

\section{Experimental Procedures}

\subsection{Laser system and optical set-up}

The lasers and the optical set-up used to realize the HD-TG experiment are
reported in fig.\ref{set-up}. The pump infrared pulses at 1064 $nm$
wavelength, are produced by a mode-locked Nd-YAG laser (Antares-Coherent)
and then they are amplified by a regenerative Nd-YAG cavity ($R3800$-Spectra
Physics) to reach 1 $mJ$ at 1 $kHZ$ of repetition rate with 100 $ps$ of
duration. The probing beam, at 532 $nm$ wavelength, is produced by a
diode-pumped intracavity-doubled Nd-YVO (Verdi-Coherent), this is a CW
single-mode laser characterized by an excellent intensity stability with low
and flat noise-intensity spectrum. The beam intensities and polarizations
are controlled by two couples of half-wave plate and polarizer.

The optical set-up, shown in fig.\ref{set-up}, uses a phase grating as a
diffractive optical element (DOE): controlling the depth of the grooves and
their spacing it is possible to obtain very high diffraction efficiency also
better than $80\%$ on the first two orders. Since we have to diffract both
1064 and 532 $nm$ a compromise must be used. The chosen DOE (made by
Edinburgh Microoptics) gives on a single beam at first order a $12\%$
diffraction efficiency for the 532 $nm$ and $38\%$ for the 1064 $nm.$
Different spacings can be used to change the $q$ vector. With the aid of a
dicroic mirror (DM), the excitation and probing beams are sent colinearly on
the DOE that produces the two excitation pulses ($E_{ex}$), the probing ($%
E_{pr}$) and the reference beam ($E_{l}$). These beams are collected by a
first achromatic lens\ (AL1), cleaned by a spatial mask to block other
diffracted orders and then recombined and focused by a second lens (AL2) on
the sample. The local laser field is also attenuated by a neutral density
filter and adjusted in phase passing through a couple of quartz slab
properly etched. The excitation grating produced on the sample is mirror
image of the enlighted DOE phase pattern. If AL1 and AL2 have the same focal
length, the excitation grating has half the spacing of DOE\cite{Maznev98}.
This type of set-up gives automatically the Bragg condition on all the beams
and produces a very stable phase locking between the probing and reference
beam, a crucial parameter to realize a heterodyne detection. To properly
test the acoustic damping \cite{Nelson} the excitation beam is focalized by
a cylindrical lens (CL2) on the DOE and the produced grating on the sample
is extended in the $q$ direction (about $5\,mm$), viceversa the probing beam
is focalized in a circular spot ($0.5\,mm$) on the sample, through the two
lens CL1 and CL2. We reduced the lasers energy on the sample to the possible
lowest level to avoid undesirable thermal effects, and the CW beams has been
gated in a window of about $1$ $ms$ every $10$ $ms$ by using a mechanical
chopper synchronized with the excitation pulses. The mean exciting energy
was 7 $mW$ ($35\,\mu J$ per pulse at \thinspace $100\,Hz$) and the probing
energy was $6$ $mW$. The reference beam intensity is very low and it is
experimentally adjusted, using a variable neutral filter, to be about 100
times the intensity of the diffracted signal. With these intensities the
experiment is deeply inside the linear response regime and no dependence of
HD-TG signal shape on the intensities of the beams can be detected. The
HD-TG signal, after it has been optically filtered, is measured by the a
fast avalanche photodiode with a bandwidth of $1$ $GHz$ (APD, Hamamatsu),
amplified and recorded by a digital oscilloscope with $1$ $GHz-4$ $Gs$
(LeCroy).

The OTP, from Fluka (99 \%), has been purified by repeated crystallization
in methanol and dried under vacuum. The sample is kept in an aluminium cell
with a teflon coating and it shows a stable supercooled phase. The cell
temperature is controlled by a cryostat system (helium closed circle,
Cryotip) with a platinum resistance dipped in the sample.

\subsection{Data collection and handling}

The OTP measured signal spans over many decades in time, typically up to
about $1$ $ms$, so we recorded the data in a pseudo-logarithmic time scale.
We use a fast time window ($0-1$ $\mu s$ range with a $250$ $ps$ time-step),
an intermediate ($0-20$ $\mu s$ range with a $10$ $ns$ time-step) and long
one ($0-1$ $ms$ range with a $200$ $ns$ time-step) and then the measurements
are merged in a single data file, we did not have any problems of
overlapping. Each data is an average of $5000$ recording (corresponding to
about 1 minute of acquisition time) and this is enough to produce an
excellent signal to noise ratio. In order to get rid of the homodyne and
other spurious signals, we recorded two measurements at different phases of
the local field, the first signal, $S_{+}$, with $\Delta \varphi _{+}=2n\pi $
and a second one, $S_{-}$, with $\Delta \varphi _{-}=2(n+1)\pi $ (see eq.\ref
{Signal}) and then we subtracted $S_{-}$ from $S_{+}$ to extract the pure HD
signal, $S_{HD}$ (see eq.\ref{SignalHD1}). From fig.\ref{HDdata} where we
have reported two typical $S_{+}$ and $S_{-}$ signals and the extracted $%
S_{HD}$, it is clear how this procedure increase substantially the quality
of data.

We measured the relaxation processes of OTP as a function of temperature for
three different value of $q=0.338$, $0.630$, $1.000$ $\mu m^{-1}$. The
wave-vectors are evaluated by the geometry of the experiment and are
affected by $0.8$ $\%$, $0.6$ $\%$ and $0.4$ $\%$ error, respectively. For
each wave-vector we take data as function of temperature in a range, $%
243-373\,K$, that largely covers the liquid and supercooled region around $%
T_{c}\sim 290$, being $T_{g}=244$ $K$ and $T_{m}=329$ $K$. In fig.\ref{data}
we show in linear-log scale some representative HD-TG data on glass-forming
OTP.

As is evident from fig.\ref{data}, the density OTP dynamics \ is
characterized by an acoustic phonon, a structural relaxation that appears as
a rise in the signal and by the final thermal relaxation. The damping of
acoustic phonon and the structural processes are strongly temperature
dependent. There is a temperature region where these 'modes' (acoustic
phonon, structural and thermal relaxation) are characterized by separated
time scales. Some physical quantities, like the sound velocity, can be
directly got from HD-TG data but a complete analysis of them requires a
fitting procedure.

We tested the response function defined in eq.s \ref{ResponseISBS}, \ref
{Responsebeta} performing a non-linear least square fit on our OTP data. The
fitting includes a convolution with the instrumental response (APD,
amplifier and oscilloscope) especially important at small times in the
acoustic oscillating part of the signal. We found this response function
able to reproduce our HD-TG data on large time windows but not on the whole
measured time scale, see fig.\ref{Fit}. Indeed the very fast time scale ( $%
0-2\;ns$) is fairly reproduced, even if an extreme care has been taken to
include all the possible signal contributions, e.g. the instantaneous signal
due to the molecular hyperpolarizzability. Actually it is not surprising
that an hydrodynamics approach is not appropriate on a fast time scale since
it does not take into account any molecular proprieties. Again, as we will
discuss later in detailed see sec.V, we found that in the long time scale ($%
0.1-1$ $ms$) and in the intermediate temperature range this response
function is not able to reproduce our data that show a relaxation pattern
that can never be explained by the used response function, see fig.\ref{slow}%
. From our fit on OTP it is clear the presence of both the excitation
mechanism: the electrostrictive effects (ISBS contribution) and the thermal
effects (ISTS). We can estimate about 60\% of ISBS against 40 \% of ISTS for 
$q=0.338$ $\mu m^{-1}$ and this ratio increases when the value of the
wave-vector increases. Nevertheless from the fitting the two contributions
can be safely disentangled, thank the linear access to the response
function. To evaluate the errors for such complex data and fitting function
is not a trivial task, we used an 'a posteriori' procedure. We repeated the
experiment getting for each temperature and wave-vector several HD-TG
signals, then we fit all the signals producing a distribution of parameters.

\section{Results}

We used the previously defined response function (see eq.s \ref{Response}, 
\ref{ResponseISBS}, \ref{Responsebeta}) to extract the information about OTP
dynamics from HD-TG data. The fitting parameters are: the acoustic frequency
and damping rate ($\omega _{A}$ and $\Gamma _{A}$), the structural
relaxation time and stretching parameter ($\tau _{S}$ and $\beta $), the
thermal relaxation time ($\tau _{H}$) and the amplitude constants ($A$, $B$
and $C$). In fig.\ref{acoustic} we report the sound velocity, $c_{A}=\omega
_{A}/q$, and the damping rate, $\Gamma _{A}$, see table \ref{TableI}. At all
the investigated temperatures and wave-vectors these two parameters are
extracted with very small uncertainties, less then 1\%. The sound velocity,
see fig.\ref{acoustic}a), shows the typical temperature dependence of
viscoelastic liquids. The velocity increases lowering the temperature and
shifts from two linear dependence regime, at high temperatures the velocity
almost corresponds to the adiabatic sound velocity, $c_{0},$ while at low
temperatures corresponds to the solid-like or 'infinite' frequency sound
velocity, $c_{\infty }$. In the transition region, starting form a few ten
of degree above $T_{c}$, there is a rapid increase of $c_{A}$ toward $%
c_{\infty }$. At higher wave-vector the shift between the two regime appears
at higher temperature. This behavior reflects the rapid variation of the
structural relaxation time with the temperature. In fact at high
temperatures the structural relaxation time is shorter than the phonon
oscillation period, $\tau _{S}<<\left( \omega _{A}\right) ^{-1}$ and the
relaxation process is coupled weakly with the sound yielding a soft damping.
Again at low temperatures when $\tau _{S}>>\left( \omega _{A}\right) ^{-1}$
the two processes are decoupled yielding again a soft damping of the sound
waves. Viceversa when $\tau _{S}\,\omega _{A}\sim 1$ the structural and
acoustic phenomena have the maximum coupling yielding a maximum in the
damping rate, see fig.\ref{acoustic}b). The dependence of $\omega _{A}$ on $q
$ produces a $q$-dependence of the coupling and also a departure $\Gamma
_{A}\varpropto q^{2}$, as is seen clearly in fig.\ref{acoustic} . $\Gamma
_{A}$ at high temperature, far from the structural region, is not far from
being proportional to $q^{2}$. In the hypothesis of a single relaxation time
we can extract its value from the damping maximum, i.e. $\tau _{S}\sim
\left( \omega _{A}\right) ^{-1}$. We found: $\tau _{S}\sim 0.6$ $nsec$ for $%
q=1$ $\mu m^{-1}$, $\tau _{S}\sim 0.9$ $nsec$ for $q=0.63$ $\mu m^{-1}$and $%
\tau _{S}\sim 1.6$ $nsec$ for $q=0.338$ $\mu m^{-1}$. This non negligible $q$%
-dependence of the structural relaxation time is again a mark of the
inadequacy of the single time simple analysis and/or a mark of the presence
of heterogeneities.

When the temperature is around $T_{c}$, the structural relaxation process is
appearing in the HD-TG pattern as a bump after the vanishing of the acoustic
oscillations, see fig. \ref{data}. In that case the structural time can
easily be extracted with confidence from the fitting procedure. Indeed only
in a $q$-dependent restricted range of temperature it has been possible to
get reliable structural parameters, $\tau _{S}$ and $\beta $, from our HD-TG
data on OTP, namely in the range $T=279-297$ $K$ for $q=1$ $\mu m^{-1}$, $%
T=277-293$ $K$ for $q=0.63$ $\mu m^{-1}$ and $T=272-287$ $K$ for $q=0.338$ $%
\mu m^{-1}$. This reflects the limitation of the fitting formulas, (eq.s \ref
{ResponseISBS}, \ref{Responsebeta}), since they fully applies only when a
time scale separation exists among the various characteristic times, as
already pointed out. Nevertheless these measurements are covering a key
temperature range around the OTP critical temperature in a ($q,t$) or ($%
q,\omega $) space quite difficult to reach by other techniques. The OTP
structural parameters are reported in fig.\ref{structural}. They are
covering three decades in time that was not previously investigated, from
about $10$ $nsec$ up to $20$ $\mu sec$. Within the uncertainties resulting
essentially from the fitting procedure, the structural relaxation times does
not show any $q$-dependence and in the stretching parameter we cannot
recognize any temperature dependent too. We want to remark that the
stretching parameter is really affected by large uncertainties, see fig.\ref
{structural}b). Unfortunately, also an other relevant quantity, i.e. the
non-ergodicity parameter $f_{q\rightarrow 0}$ which can be estimated from
eq. \ref{DebeyWaller}, is affected by large uncertainties. Within them, the $%
f_{q\rightarrow 0}$ parameter of OTP from our HD-TG data do not show the
cusp-like behavior predicted by the mode-coupling theory\cite{Goetze92}, as
is evident from fig.\ref{DWfig}. On the other hand, the behavior of the
non-ergodicity parameter has been measured by neutron scattering for
different high value of wave-vectors and by extrapolating its slope in the
limit $q\rightarrow 0$ the cusp-like seem to be hardly visible \cite{Wuttke}.

The fit parameter $\tau _{H}$, the thermal relaxation time, defines the
final decay of the HD-TG signal and is safely extracted when the condition $%
\tau _{H}\gg \tau _{S}$ is verified. Since $\tau _{H}$ $\sim $ 100 $\mu $sec
and it is typically not strongly temperature dependent until $Tg$\cite
{Nelson}, it should be safely extracted in the range from $373$ $K$ down to $%
270$ $K$. In figure \ref{thermal} we report the thermal diffusivities, $\chi
=\left( \tau _{H}q^{2}\right) ^{-1}={\em k}(\rho _{o}C_{p})^{-1}$, resulting
from the investigated three wave-vectors in the\ whole temperature range. In
the high temperature range $\ \chi $ shows the expected smooth variation and
independence of $q$. Approaching the critical temperature, a strong
deviation is showing up: starting from about $275$ $K$ at $q=0.338$ $\mu
m^{-1}$ ( and from temperatures even higher at higher $q$) an anomalous
strong decrease in the thermal diffusivity is coming from our fit. The
decrease is so strong that the longer decay is clearly visible if we plot
the data in a semilogarithmic scale, as in fig.\ref{slow}. This effect could
be an interaction between the structural and thermal relaxation time
constant, $\tau _{S}$ and $\tau _{H},$ that are getting closer decreasing
the temperature. However, when the thermal diffusivity starts to deviate,
the structural and thermal relaxation times \ seem to be too far each other,
e.g. at $q=0.338$ $\mu m^{-1}$ the deviation start at about $T=275$ $K$,
where $\tau _{S}=15$ $\mu s$ and $\tau _{H}=115$ $\mu s$. Anyway we want to
remember that $\tau _{S}$ is a parameter of a time distribution rather than
the real relaxation time. Further, in the region of the thermal diffusivity
dip, the HD-TG data show at long times a second very slow decay: indeed it
is clear from fig.\ref{slow}, that the slow relaxation at $T=283$ $K$ and $%
253$ $K$ are characterized by a single exponential relaxation, the thermal
decay, and viceversa at $267$ $K$ some extra relaxation is appearing. It is
fairly obvious that the used response function (eq.s \ref{ResponseISBS}, \ref
{Responsebeta}) is not able at all to reproduce these features. As regards
the subsequent large increase of the thermal diffusivity at low temperature
where the structural relaxation times are found even longer than the thermal
relaxation, the experiments of specific heat spectroscopy on glass-formers
\cite{Birge86} suggest that the increase can be due to $C_{p}$ approaching
the solid-like or high frequency response.

\section{Discussion and conclusions}

The present analysis in terms of density hydrodynamics response of \ the
HD-TG data of glass-forming OTP around the critical temperature of the
structural arrest, shows some clear results but also open some problems in
the data interpretation. To get a further understanding, we compare our
parameters with the other data from literature. In fig.\ref{velocity} we
report our sound velocities together with the data from Brilluoin LS \cite
{Monaco2001} and ultrasound experiments\cite{Arrigo}. Here the agreement is
very nice. As expected, there is a clear dispersion effect in $q$, as the LS
data are taken at much higher wave-vectors, and the extreme values $c_{0}$
and $c_{\infty }$ are approached at high and low temperatures respectively.
It is evident that the HD-TG experiment is able to measure the sound
velocity \ at wave-vectors in an otherwise difficult range, too low for LS
techniques but too high for ultrasound experiments. In fact, in our
experiment $\omega _{A}$ ranges from $0.4$ to $2$ $GHz$ with $\Gamma
_{A}\sim 2-160$ $MHz.$

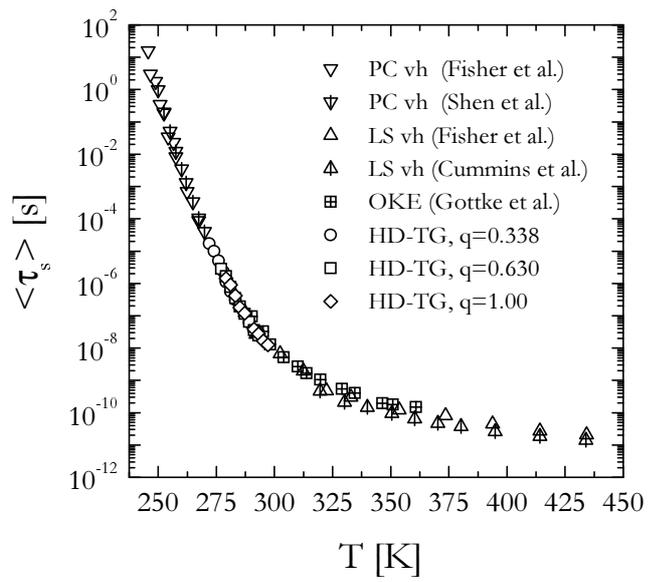
\begin{figure}[tbp]
\caption{OTP structural relaxation times from HD-TG data, LS \protect\cite
{Fisher94,Cummins97},  PC \protect\cite{Fisher94,Shen99}, and OKE data 
\protect\cite{Fayer2001}.}
\label{taus}
\end{figure}

Our structural relaxation times are compared with the light scattering (LS) 
\cite{Fisher94,Cummins97}, photon correlation (PC) \cite{Fisher94,Shen99}
and time-resolved optical Kerr effect (OKE) data \cite{Fayer2001}. All these
data are extracted from experiments performed with depolarized light
geometry and so the influence of the orientational dynamics can be non
negligible. Nevertheless, our OTP data, spanning from $10^{-8}$ up to $%
10^{-5}$ sec, nicely cover the gap present between LS, OKE and PC
measurements and they follow the same course, as we can see in fig. \ref
{taus}. This good agreement has some implication in OTP dynamics suggesting
that the orientational and translational dynamics have substantially the
same temperature dependence, in the investigated temperature range, with
very similar relaxation times.

The cusp-like behavior of the Debye-Waller factor expected in the frame of
mode-coupling theory for a fragile glass such as OTP, is not recognizable,
owing to large uncertainties.

We would like to stress that the slow thermal decay, at relatively low
temperature, did not show a simple single exponential decay and can not be
properly characterized by the simple hydrodynamics model used in this work.
We have found this behavior common to other several glass-formers: glycerol, 
{\it m}-toluidine, salol \cite{Sampoli}. In our opinion the apparent
modification of the thermal diffusivity has to be addressed to effects of
the structural dynamics on the thermalization process. In other word, the
heat flux, produced by the weak absorption of the pump laser pulses, modify
the kinetic energy of the different degree of freedom and not all of them
thermalize in a very short time scale. Furthermore part of the
roto-translational energy thermalizes through collective rearragements that
act on the structural time scale. This kind of processes are also
responsible of the frequency dependence of the specific heat. A more
suitable hydrodynamic treatment for the heat transport should be able to
taken into account the complex relaxation patterns found in the present
experiment.

\begin{acknowledgements}
We thank F.Barocchi, K.A.Nelson, R.M.Pick and G.Ruocco for very helpful 
suggestions and discussions. This work was supported by the Commission of the European
Communities through the contract No HPRI-CT1999-00111, by MURSTand by INFM through the project TREB-Sez.C-PAISS1999.
\end{acknowledgements}

\begin{table} [tbp] \centering
\renewcommand{\baselinestretch}{1}
%
\begin{tabular}{ccccccccc}
& \multicolumn{2}{c}{$q=0.338$ $\mu m^{-1}$} & \multicolumn{2}{c}{$q=0.630$ $%
\mu m^{-1}$} & \multicolumn{2}{c}{$q=1.00$ $\mu m^{-1}$} &  &  \\ 
$T$ & $\omega _{A}$ & $\Gamma _{A}$ & $\omega _{A}$ & $\Gamma _{A}$ & $%
\omega _{A}$ & $\Gamma _{A}$ & $\frac{\Delta \omega _{A}}{\omega _{A}}$ & $%
\frac{\Delta \Gamma _{A}}{\Gamma _{A}}$ \\ 
$(K)$ & $(GHz)$ & $(MHz)$ & $(GHz)$ & $(MHz)$ & $(GHz)$ & $(MHz)$ &  &  \\ 
\hline
243.2 & -- & -- & 1.572 & 3.30 & -- & -- & $\eqslantless 0.1$ \% & $%
\eqslantless 1$ \% \\ 
244 & 0.8348 & 2.94 & -- & -- & -- & -- & '' & '' \\ 
253.2 & -- & -- & 1.525 & 3.81 & 2.423 & 6.47 & '' & '' \\ 
254 & 0.8094 & 3.02 & -- & -- & -- & -- & '' & '' \\ 
263.2 & -- & -- & 1.474 & 6.90 & 2.343 & 8.57 & '' & '' \\ 
264 & 0.7821 & 3.52 & -- & -- & -- & -- & '' & '' \\ 
270 & 0.7629 & 4.53 & -- & -- & -- & -- & '' & '' \\ 
271.2 & -- & -- & 1.424 & 8.90 & 2.273 & 12.7 & '' & '' \\ 
279.2 & 0.7275 & 10.1 & 1.381 & 15.2 & 2.195 & 22.3 & '' & '' \\ 
287.2 & 0.693 & 22.1 & 1.319 & 35 & 2.10 & 45 & $\eqslantless 0.5$ \% & $%
\eqslantless 5$ \% \\ 
295.2 & 0.642 & 44 & 1.240 & 68 & 2.00 & 86 & '' & '' \\ 
303.2 & -- & -- & 1.139 & 93 & 1.868 & 142 & '' & '' \\ 
304 & 0.577 & 50 & -- & -- & -- & -- & '' & '' \\ 
313.2 & -- & -- & 1.025 & 86 & 1.674 & 156 & '' & $"$ \\ 
314 & 0.52283 & 30 & -- & -- & -- & -- & '' & '' \\ 
323.2 & -- & -- & 0.948 & 41 & 1.539 & 95 & $\eqslantless 0.1$ \% & $%
\eqslantless 1$ \% \\ 
324 & 0.4977 & 11.9 & -- & -- & -- & -- & '' & '' \\ 
333.2 & -- & -- & 0.913 & 20.8 & 1.462 & 52.7 & '' & '' \\ 
334 & 0.4837 & 5.96 & -- & -- & -- & -- & '' & '' \\ 
343.2 & -- & -- & 0.888 & 12.8 & 1.415 & 31.5 & '' & '' \\ 
344 & 0.4715 & 3.87 & -- & -- & -- & -- & '' & '' \\ 
353.2 & -- & -- & 0.866 & 9.37 & 1.377 & 21.8 & '' & '' \\ 
354 & 0.4600 & 2.93 & -- & -- & -- & -- & '' & '' \\ 
363.2 & -- & -- & 0.8441 & 7.34 & 1.342 & 15.8 & '' & '' \\ 
364 & 0.4485 & 2.47 & -- & -- & -- & -- & '' & '' \\ 
373.2 & -- & -- & 0.8236 & 6.38 & 1.309 & 12.8 & '' & '' \\ 
374 & 0.4375 & 2.23 & -- & -- & -- & -- & '' & ''
\end{tabular}
\caption{Sound frequencies and damping rates with  their relative errors 
for the analysed q-values in the range 243 - 374. To avoid overcrowded table we report only some temperatures.
K.\label{TableI}}%
\end{table}%
%

\begin{figure}[tbp]
\caption{Optical set-up and laser system for HD-TG experiment with optical
heterodyne detection: M: mirror; CL\#: cilyndrical lens; DM: dicroic mirror;
DOE: diffractive optic element; AL\#: achromatic lens; APD: avalanche
photodiode.}
\label{set-up}
\end{figure}

\begin{figure}[tbp]
\caption{Typical HD-TG raw data corresponding to phase difference between
signal and local reference of $\Delta \protect\varphi =0%
{{}^\circ}%
$ and $\Delta \protect\varphi =180%
{{}^\circ}%
$ (above); pure HD-TG signal (below) is obtained by substracting the two
previous signals. The subtraction permits to improve substantially the
signal/noise ratio and to remove the homodyne and all spurious contributions
that are phase independent.}
\label{HDdata}
\end{figure}

\begin{figure}[tbp]
\caption{HD-TG data on OTP for two wave-vectors, $q=0.338$ and $q=1$ $%
\protect\mu m^{-1}$ , at several temperatures. The data, at all
temperatures, show damped acoustic oscillations at short times and thermal
diffusion at long times. Decreasing of temperature the structural relaxation
mode, typical of complex liquids, appears at first as strong acoustic
damping ($\protect\omega _{A}\protect\tau _{S}\thicksim 1$ condition) and
later as gradual rise of TG signal.}
\label{data}
\end{figure}

\begin{figure}[tbp]
\caption{HD-TG data (solid lines), fits (dotted lines) and residues x3
(lower lines) for OTP, at the two temperatures, $T=288K$ and $T=267K$ at
wave-vector $q=0.338$ $\protect\mu m^{-1}$.}
\label{Fit}
\end{figure}

\begin{figure}[tbp]
\caption{Sound velocities $C_{A}$ (a) and acoustic damping rates $\Gamma _{A}
$ (b) versus temperature from fits to HD-TG data at the investigated
wave-vectors. At each q-value the sound dispersion and damping rate reach
the maximum at the temperature at which $\protect\omega _{A}\protect\tau
_{S}\sim 1.$}
\label{acoustic}
\end{figure}

\begin{figure}[tbp]
\caption[Structural relaxation times $\protect\tau _{R}$ and stretching
parameters $\protect\beta $ versus temperatures at the three wave-vectors
analysed.]{Structural relaxation times $\protect\tau _{S}$ and stretching
parameter $\protect\beta $ versus temperature at the three wave-vectors
analysed.}
\label{structural}
\end{figure}

\begin{figure}[tbp]
\caption{Temperature dependence of the Debye-Waller factor, $f_{q\rightarrow
0}(T)$, in OTP, obtained by fits, at all wave-vectors. There is no evidence
of a MCT cusp in the analysed temperature range .}
\label{DWfig}
\end{figure}

\begin{figure}[tbp]
\caption{Thermal diffusivity versus temperature. For every investigated
q-value, the extracted thermal diffusivity show, at low temperatures, a
peculiar behavior not in agreement with the temperature dependence of the
thermodynamic thermal diffusivity.}
\label{thermal}
\end{figure}

\begin{figure}[tbp]
\caption{HD-TG data (solid curves) and fits (dotted curves), at wave-vector $%
q=0.338$ $\protect\mu m^{-1}$ at some temperatures, are plotted in log-lin
scale to show the complex decay at long times. In the intermediate
temperatures range, where the thermal diffusivity (fig. 8) shows the
anomalous dip, the used fitting function (\ref{Responsebeta}), is not able
to reproduce the tail at long times.}
\label{slow}
\end{figure}

\begin{figure}[tbp]
\caption{OTP sound velocities from HD-TG data, LS data \protect\cite
{Monaco2001} and ultrasonic measurements \protect\cite{Arrigo}.}
\label{velocity}
\end{figure}

\begin{references}
\bibitem{ISBSstructural}  According to the hydrodynamics model \cite{Nelson}
the whole ISBS signal is the following: 
\[
G_{\rho \rho }(q,t)\simeq C\left[ e^{-\Gamma _{A}t}\sin \left( \omega
_{A}t\right) \right] +D\left[ -e^{-\Gamma _{A}t}\cos (\omega
_{A}t)+e^{-(t/\tau _{S})^{\beta }}\right] 
\]
Indeed the amplitude $D$ has been found in the previous work to be
negligible. Also in the present work we found the $D$ terms to be very small
and not affecting at all the fitting results Nevertheless, since this test
is based only on the fitting procedure not a final conclusion can be given
about the intensity of structural contribution in the ISBS signal.
\end{references}
\end{document}